# Shaping soliton properties in optical Mathieu lattices

Yaroslav V. Kartashov, Alexey A. Egorov,* Victor A. Vysloukh,** and Lluis Torner

*ICFO-Institut de Ciencies Fotoniques, and Universitat Politecnica de Catalunya,*

*08034, Barcelona, Spain*

We address basic properties and stability of two-dimensional solitons in photonic lattices induced by the nondiffracting Mathieu beams. Such lattices allow for smooth topological transformation of radially symmetric Bessel lattices into quasi-one-dimensional periodic ones. The transformation of lattice topology drastically affects properties of ground-state and dipole-mode solitons, including their shape, stability and transverse mobility.

Transverse variations of refractive index of nonlinear media drastically alter the propagation of optical solitons, which can henceforth be routed and steered. The technique of optical induction, recently introduced in nonlinear optics [1-5], opens broad horizons for creation of various transverse refractive index landscapes, or optical lattices. Lattices with tunable features can be thus readily imprinted in suitable crystals. The basic properties of solitons supported by the lattice are defined by its topology. Thus, domain of soliton existence in simplest periodic lattice, formed by set of plane waves, is dictated by the Floquet-Bloch lattice spectrum [1-10]. Another important class of optical lattices can be created by nondiffracting Bessel beams with radial symmetry. Such lattice symmetry results in new soliton features and opens new ways for soliton manipulation, including the possibility to induce rotary soliton motions and collisions in different lattice rings [11-13], as well as reconfigurable soliton networks [14,15].

In this Letter we study a new type of optical lattice that sets important connection between periodic and radially symmetric Bessel lattices. Such lattices can be induced by nondiffracting Mathieu beams, and afford smooth topological transformation of radially symmetric profile of Bessel lattice into quasi-one-dimensional periodic one. The transformation of the lattice topology finds its manifestation in dramatic change of properties of ground-state and dipole-mode solitons.

We start our analysis with a generic equation describing the propagation of laser radiation along $\xi$ axis of Kerr-type cubic nonlinear medium with imprinted transverse modulation of refractive index

$$i\frac{\partial q}{\partial \xi} = -\frac{1}{2}\left(\frac{\partial^2 q}{\partial \eta^2} + \frac{\partial^2 q}{\partial \zeta^2}\right) + \sigma q |q|^2 - pR(\eta,\zeta)q. \qquad (1)$$

Here $q(\eta,\zeta,\xi)$ is the dimensionless amplitude of the light field; the longitudinal $\xi$ and transverse $\eta,\zeta$ coordinates are scaled to the diffraction length and the input beam width respectively; parameter $\sigma = \pm 1$ stands for the nonlinearity sign (defocusing/focusing); $p$ describes the lattice depth and the function $R(\eta,\zeta)$ stands for the transverse lattice profile. We assume that optical lattice features intensity $R(\eta,\zeta) \sim |q_\mathrm{M}|^2$ of nondiffracting Mathieu beam, similarly to periodic lattices created in a photorefractive medium [1-10], while $\max[R(\eta,\zeta)] = 1$. Field distribution of Mathieu beam can be written via Whittaker integral [16-18]:

$$q_\mathrm{M}(\eta,\zeta,\xi) = \exp(ib_\mathrm{lin}\xi)\int_{-\infty}^{\infty} A(\phi)\exp[i(-2b_\mathrm{lin})^{1/2}(\eta\cos\phi + \zeta\sin\phi)]d\phi, \qquad (2)$$

where $b_\mathrm{lin} < 0$ is the propagation constant; $A(\phi)$ is the angular spectrum represented by even $\mathrm{ce}_m(\phi,-b_\mathrm{lin}a^2/2)$, $m = 0,1,2...$, or odd $\mathrm{se}_m(\phi,-b_\mathrm{lin}a^2/2)$, $m = 1,2,3...$ angular Mathieu functions, and $a$ is the interfocal parameter. Mathieu beams are fundamental nondiffracting solutions of the wave equation in elliptic cylindrical coordinates. They can be generated by illumination of narrow annular slit with Gaussian aperture placed in the focal plane of a lens [17,18], while the beam topology can be controlled by the width of aperture. Eq. (1) conserves the energy flow $U = \iint_{-\infty}^{\infty}|q|^2\,d\eta d\zeta$.

Representative examples of lowest order even ($m = 0$) and odd ($m = 1$) Mathieu lattices that we consider in this Letter are shown in Fig. 1. For small values of interfocal parameter $a \to 0$ the foci of associated elliptical coordinate system collapse to a point, and even Mathieu lattice transforms into radially symmetric Bessel lattice. Odd Mathieu beams produce azimuthally modulated lattices at $a \to 0$. At $a \to \infty$ when separation of the foci tends to infinity, lattice transforms into quasi-one-dimensional periodic pattern. Thus, modification of interfocal parameter results in a smooth topological deformation of lattice shape. Notice, that lattice frequency in $\zeta$-direction is dictated by the parameter $b_{\text{lin}}$, so further we fix $b_{\text{lin}} = -2$ and vary $a$ and $p$.

Profiles of simplest solitons supported by Mathieu lattices are depicted in Fig. 2. We found them in the form $q(\eta,\zeta,\xi) = w(\eta,\zeta)\exp(ib\xi)$, where $w(\eta,\zeta)$ is the real function and $b$ is the propagation constant. Ground-state solitons reside on the central maximum of simplest ($m = 0$) even Mathieu lattice imprinted in focusing medium. At $a = 0$ such solitons are radially symmetric. With increase of $a$ at fixed $U$ and $p$ solitons gradually become elliptic, their amplitude decreases, and solitons expand over neighboring lattice maxima along $\zeta$-axis. For large $a$ values solitons may feature strong modulation along $\zeta$-axis, whose depth increases with increase of lattice depth. Mathieu lattices imprinted in focusing medium support dipole-mode solitons whose field changes sign between two central lattice maximums (below we consider such solitons in odd lattices with $m = 1$). Lattices in defocusing medium support specific solitons that can be termed gap solitons. Localization mechanism for such solitons in $\zeta$-direction (where lattice is almost periodic for $a \gg 1$) is of Bragg type, while confinement in $\eta$ is achieved because of appropriate refractive compensation of diffraction and defocusing. In-phase combinations of such gap solitons are also possible (Fig. 2).

Transformation of lattice topology substantially affects solitons stability. Properties of ground-state solitons at $\sigma = -1$ are summarized in Fig. 3. At small values of $a$, the energy flow monotonically increases with $b$ (Fig. 3(a)). Energy flow vanishes in the cutoff on $b$ that increases with lattice depth $p$ and interfocal parameter $a$ (Fig. 3(b)). With increase of $b$, solitons get narrower, their amplitude increases, and their energy flow asymptotically approaches $U_{\text{cr}} \approx 5.85$ that corresponds to energy flow of Townes soliton in uniform medium. Thus, at $a = 50$, $p = 1$, and $b = 2$ soliton peak amplitude $\approx 2.78$ and $U \approx 5.23$, while at $b = 7.5$ peak amplitude reaches 6.22 and $U \approx 5.69$. At a given critical value of $a$, a branch with negative $dU/db$ appears. According to the Vakhitov-Kolokolov (VK) criterion this branch corresponds to unstable solitons. Solitons belonging to this branch expand over many lattice periods in $\zeta$ direction (Fig. 2) that results in increase of $U$ with decrease of $b$, so that energy flow becomes a two-valued function of $b$. Such soliton expansion in the small amplitude limit is consistent with the fact that linear guided modes of Mathieu lattices for large $a$ approach delocalized Bloch modes of quasi-one-dimensional periodic patterns. In contrast, at $a \ll 1$ low-amplitude solitons transform into linear guided modes of almost radially symmetric lattice, which are always well localized. The instability domain for ground-state soliton broadens with $a$ (Fig. 3(c)). Direct stability analysis performed with Eq. (1), linearized around perturbed stationary solution $w(\eta,\zeta)$, confirmed conclusions based on the VK criterion. Instability of ground-state soliton in Mathieu lattice at $a \gg 1$ is associated with exponentially growing perturbation (Fig. 3(d)).

The topology of Mathieu lattice has even stronger impact on properties of dipole-mode solitons. The energy flow of such solitons increases monotonically with $b$ at small $a$ (Fig. 4(a)). Energy flow vanishes in the cutoff, where solitons broadens substantially. For high enough $a$ soliton ceases to exist in the cutoff without any topological shape transformation. This behavior finds its manifestation in the discontinuity of dependence of cutoff on interfocal parameter (Fig. 4(b)). With growth of $b$ dipole-mode soliton transforms into two narrow weakly interacting out-of-phase solitons. Stability analysis revealed that dipole-mode solitons become stable above a threshold on propagation constant (or $U$) similarly to their counterparts in two-dimensional periodic lattices [8] (Fig. 4(c)). Dipole-mode solitons suffer oscillatory instabilities associated with complex growth rates. Importantly, we found that the threshold value of propagation constant for stabilization quickly increases with $a$, upon transformation of lattice into quasi-one-dimensional one, so that at $a \to \infty$ dipole-mode solitons become unstable in the entire domain of their existence. This shows the drastic difference between stability properties of dipole-mode solitons in quasi-one-dimensional and two-dimensional lattices, which always supports stable dipole-mode solitons. Gap solitons and their in-phase combinations depicted in Fig. 2 can also be made stable in Mathieu lattices.

The transformation of the lattice topology with increase of $a$ substantially affects soliton mobility. Solitons are strongly pinned by almost radially symmetric lattices with small $a$, and can hardly jump into neighboring lattice rings. In contrast, quasi-one-dimensional lattices with large $a$ feature pronounced channels along $\eta$-axis, so that even small input tilts cause considerable transverse displacements (periodic oscillations) of both ground-state and dipole-mode solitons (Fig. 4(d)). The amplitude of these oscillations increases with $a$ for a given input tilt. Notice also the possibility to induce rotary motions in lowest-order even Mathieu lattices with small interfocal parameters.

Summarizing, we analyzed properties of solitons supported by Mathieu lattices, that sets important connection between radially symmetric and quasi-one-dimensional periodic lattices. We showed how lattice topology affects basic properties of ground-state and dipole-mode solitons including their stability, and mobility in transverse plane.

*Visiting from the Physics Department, M. V. Lomonosov Moscow State University, Moscow, Russia. **Visiting from the Departamento de Fisica y Matematicas, Universidad de las Americas, Puebla, Mexico.

# References with titles

# References without titles

# Figure captions

Figure 1. Lowest-order even (top row) and odd (bottom row) Mathieu lattices with different values of interfocal parameter $a$.

Figure 2. Top row: ground-state soliton in even lattice at $b=2.9$, $a=500$, $p=5$ (left) and dipole-mode soliton in odd lattice at $b=2.84$, $a=30$, $p=5$ (right). Focusing medium. Bottom row: simplest gap soliton in even lattice at $b=3.2$, $a=12$, $p=15$ (left) and combination of in-phase gap solitons in odd lattice at $b=3.2$, $a=12$, $p=15$ (right). Defocusing medium.

Figure 3. Properties of ground-state solitons in even Mathieu lattices. (a) Energy flow versus propagation constant. (b) Propagation constant cutoff versus interfocal parameter. (c) Boundaries of instability domain versus interfocal parameter. (d) Real part of perturbation growth rate versus propagation constant. In all cases $p=1$. Focusing medium.

Figure 4. Properties of dipole-mode solitons in odd Mathieu lattices. (a) Energy flow versus propagation constant. (b) Propagation constant cutoff versus interfocal parameter. (c) Real part of perturbation growth rate versus propagation constant. (d) Trajectory of center of ground-state soliton with $b=4$ and input tilt $\alpha_\eta = 1$ launched into even lattices corresponding to different $a$. In all cases $p=5$. Focusing medium.

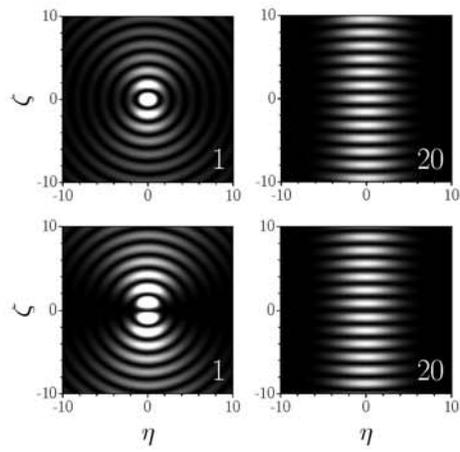

Figure 1. Lowest-order even (top row) and odd (bottom row) Mathieu lattices with different values of interfocal parameter $a$.

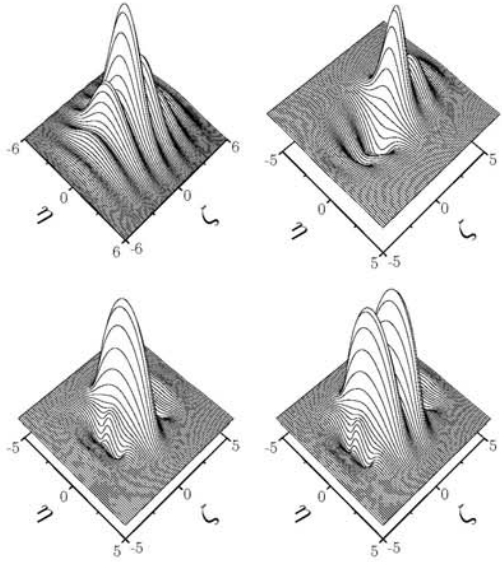

Figure 2. Top row: ground-state soliton in even lattice at $b = 2.9$, $a = 500$, $p = 5$ (left) and dipole-mode soliton in odd lattice at $b = 2.84$, $a = 30$, $p = 5$ (right). Focusing medium. Bottom row: simplest gap soliton in even lattice at $b = 3.2$, $a = 12$, $p = 15$ (left) and combination of in-phase gap solitons in odd lattice at $b = 3.2$, $a = 12$, $p = 15$ (right). Defocusing medium.

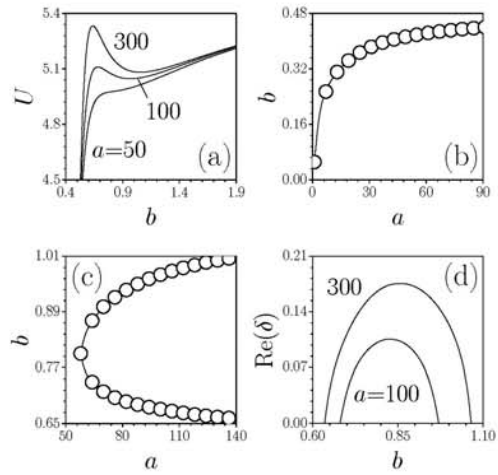

Figure 3. Properties of ground-state solitons in even Mathieu lattices. (a) Energy flow versus propagation constant. (b) Propagation constant cutoff versus interfocal parameter. (c) Boundaries of instability domain versus interfocal parameter. (d) Real part of perturbation growth rate versus propagation constant. In all cases $p=1$. Focusing medium.

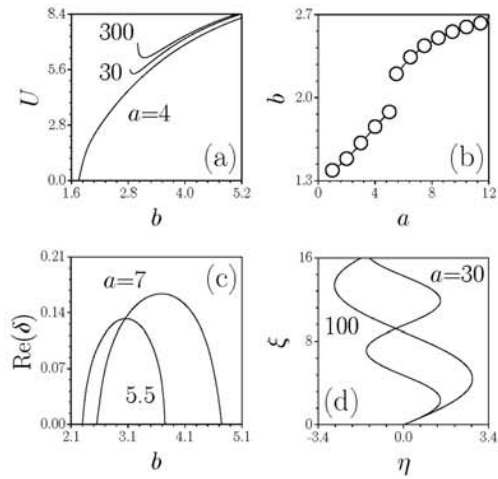

Figure 4. Properties of dipole-mode solitons in odd Mathieu lattices. (a) Energy flow versus propagation constant. (b) Propagation constant cutoff versus interfocal parameter. (c) Real part of perturbation growth rate versus propagation constant. (d) Trajectory of center of ground-state soliton with $b = 4$ and input tilt $\alpha_\eta = 1$ launched into even lattices corresponding to different $a$. In all cases $p = 5$. Focusing medium.